\documentclass[%
reprint,amsmath,amssymb,aps,prb,superscriptaddress,longbibliography,
floatfix,
]{revtex4-2}

\usepackage{graphicx}
\usepackage{dcolumn}
\usepackage{bm}
\usepackage{epstopdf}
\usepackage{float}
\usepackage{braket}
\usepackage{dsfont}
\usepackage{amsmath}
\usepackage{siunitx} 
\usepackage[italicdiff]{physics}
\usepackage[T1]{fontenc}
\usepackage[dvipsnames]{xcolor}
\usepackage{color,soul}
\usepackage{lipsum}
\usepackage[colorlinks=true, allcolors=blue]{hyperref}

\graphicspath{{figures_v5/}}

\begin{document}

\title{Measuring pulse heating in Si quantum dots with individual two-level fluctuators}

\author{Feiyang Ye}

\author{Lokendra S. Dhami}

\affiliation{Department of Physics and Astronomy, University of Rochester, Rochester, NY, 14627, USA}

\affiliation{University of Rochester Center for Coherence and Quantum Science, Rochester, NY, 14627, USA}

\author{John M. Nichol}
\email{john.nichol@rochester.edu}
\affiliation{Department of Physics and Astronomy, University of Rochester, Rochester, NY, 14627, USA}

\affiliation{University of Rochester Center for Coherence and Quantum Science, Rochester, NY, 14627, USA}

\begin{abstract}
To encode quantum information in semiconductor spin qubits, voltage pulses are necessary for initialization, gate operation, and readout. However, these pulses dissipate heat, shifting spin-qubit frequencies and reducing gate fidelities. The cause of this pulse heating in quantum-dot devices is unknown. Here, we measure pulse heating using charged two-level fluctuators (TLFs) in Si/SiGe quantum dots. Specifically, we observe that voltage pulses on nearby gates tend to increase TLF switching rates and occupation biases.  The amount of heating depends on the pulse amplitude and frequency, but not on the distance between the pulsed gates and the TLFs. The amount of heating also generally depends on the idling voltage of the pulsed gates, suggesting that electrons accumulated under or near the gates contribute to the heating. We hypothesize that reducing the area of the gates with electrons nearby could mitigate the heating.
\end{abstract}

\pacs{}

\maketitle

\section{Introduction}
Spin qubits in silicon quantum dots show promise as the building blocks of large-scale quantum computers, due to their small footprint, long coherence times, and compatibility with semiconductor technology \cite{burkard2023semiconductor}.
Most semiconductor spin qubits rely on fast and precise voltage pulses to achieve initialization, gate operations, and readout.
However, spurious effects caused by voltage pulses can complicate qubit operations \cite{freer2017single,takeda2018optimized, yoneda_quantum-dot_2018, zwerver2022qubits, philips2022universal, doi:10.1126/sciadv.add9408, PhysRevX.13.041015, tanttu2024assessment}, posing challenges for fault-tolerant quantum computing with spin qubits.

A common experimental observation is that spin qubit frequencies shift with voltage pulses \cite{takeda2018optimized, yoneda_quantum-dot_2018, philips2022universal, zwerver2022qubits, doi:10.1126/sciadv.add9408, PhysRevX.13.041015, tanttu2024assessment}, depending on the pulse duration and amplitude.
The microscopic origin of the pulse-induced frequency shift remains poorly understood, but is potentially caused by the effect of heat from the pulses~\cite{PhysRevX.13.041015} on electron $g$-factor variations \cite{doi:10.1126/sciadv.add9408, PhysRevX.13.041015} and/or charge fluctuators \cite{PhysRevResearch.6.013168, sato2024simulation}.
The origin of this pulse heating in quantum-dot devices is also unknown and remains an open question.  

Regardless of its potential link to pulse-induced frequency shifts, pulse heating is a critical challenge for semiconductor spin qubits. Excess heat in semiconductor quantum dot devices may affect not only qubit resonance frequencies but also charge noise levels, as well as initialization and readout fidelities \cite{burkard2023semiconductor}.
Given these challenges, understanding pulse heating is essential. Mesoscopic cryogenic thermometry and calorimetry techniques~\cite{Giazotto2006RMP}, including the use of hybrid junctions \cite{PhysRevApplied.3.014007,gumucs2023calorimetry}, quantum dots \cite{maradan2014gaas, nicoli2019QD, PhysRevApplied.15.034044, PhysRevApplied.21.064039}, and noise thermometry~\cite{Qu_2019,Spietz_2003} could potentially be used to study pulse heating, but each of these techniques requires additional experimental overhead.

In this work, we use naturally occurring charged two-level fluctuators (TLFs), instead of a separated dedicated thermometer, to study temperature changes in silicon quantum dots in response to pulse heating. The advantage of this technique is that it requires no extra fabrication or device complexity yet enables new insights into pulse heating in semiconductor quantum dots.  
Although their precise nature is not well understood, TLF properties can be sensitive to temperature variations \cite{lisenfeld2010measuring,wheeler2020sub,ye2024characterization}.
Indeed, regardless of the underlying microscopic model \cite{Dutta1981,phillips1987two,Pourkabirian2014,Paladino2014,muller2019towards}, TLF transition rates and occupation biases are expected to depend on temperature. 

We leverage this temperature sensitivity to study pulse heating in Si/SiGe quantum dots. We find that the heating depends on the pulse amplitude and frequency, but not on the distance between the pulsed gates and the TLFs. 
Additionally, we consistently find that the heating depends on the value of the idling voltage. In particular, the behavior differs when the gates are biased above or below the turn-on voltage, suggesting that electron accumulation under or near the gates contributes to the observed heating.
Our work also suggests potential mitigation strategies for the pulse heating.

\section{Experimental setup}
\begin{figure}[ht!]
{\includegraphics[width= 0.5\textwidth]{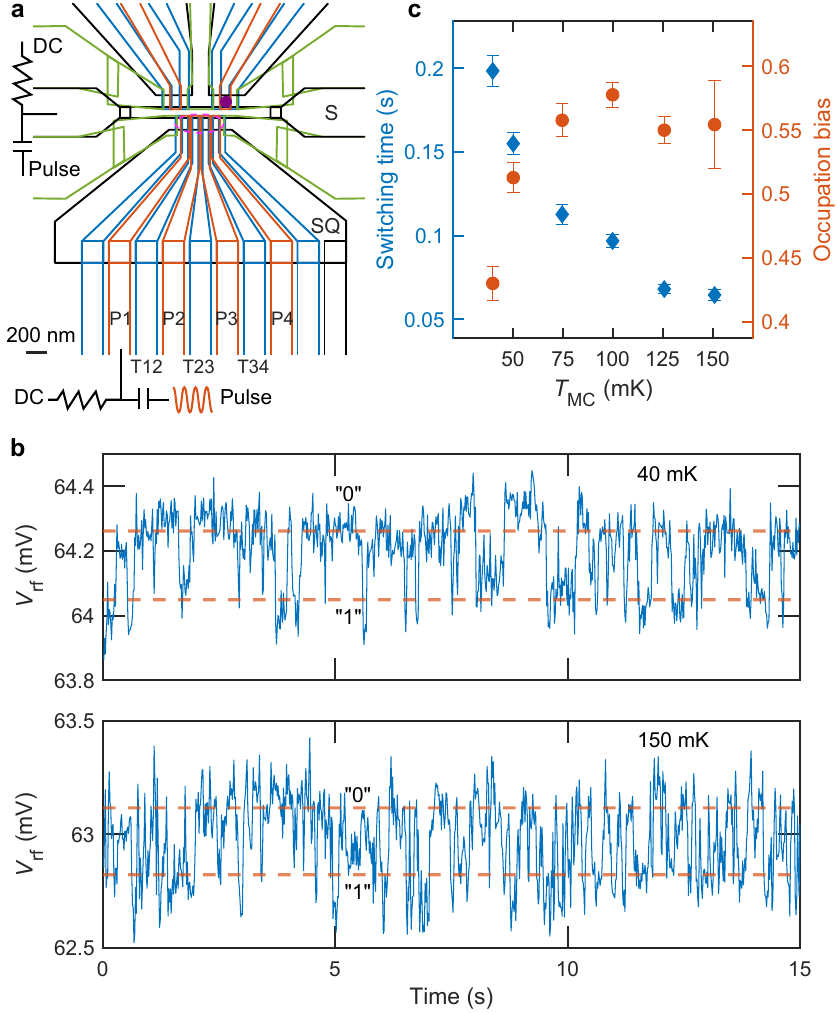}} 
\caption{
\textbf{Experimental setup.}
\textbf{a} Gate pattern for Device 1. The upper right quantum dot is used as a charge sensor to measure the TLF. Plunger gates, tunneling gates, accumulation gates, and screening gates are sketched in red, blue, green, and black, respectively. The black scale bar is $200~\si{nm}$. The lower main-side channel circled by a purple box is closed by screening gates S and SQ.
Voltage pulses are applied through the middle screening gate S and seven finger gates on the lower main side via bias tees. All the pulse amplitudes reported in this work are at the device level after attenuators and bias tees (see Ref.~\cite{ejcthesis} for further details about the experimental setup).
Electrons are occupied under the gate-stack fanout regions when idling gate voltages are above the accumulation threshold. In the fanout regions, the total area that each finger gate overlaps the quantum well is about $3,300~\si{\mu m^2}$.
\textbf{b} Examples of TLF time traces at mixing chamber temperatures of $40~\si{mK}$ and $150~\si{mK}$. The two subplots share a common $x$-axis. The horizontal red dashed lines represent the mean values associated with the  ``0'' and ``1'' states.
\textbf{c} TLF switching time and occupation bias vs. mixing chamber temperature $T_\text{MC}$. 
The switching times $\tau=t_\text{max}/0.946$ are extracted from the Allan variance of the time series \cite{ye2024characterization}, where $t_\text{max}$ is the time lag where the Allan variance has a peak.
The occupation biases $B$ are computed by fitting the signal histogram to the sum of two Gaussians with the same variance, and the error bars are calculated from the $95\%$ confidence bounds of the fit parameters. The reduced sensor sensitivity above $100~\si{mK}$ due to thermal broadening of the transport peak causes a large overlap of the two Gaussians, which could introduce errors in the estimated occupation bias at elevated temperatures.
}
\label{fig:setup}
\end{figure}

We have measured two devices in two different but similar dilution refrigerators with a base temperature of about $10~\si{mK}$. Device 1 is a quadruple-quantum-dot device with two charge sensors (Fig.~\ref{fig:setup}a), and Device 2 is a double-quantum-dot device with one charge sensor (see Supplemental Figure S6 in the Supplemental Material \cite{SMnote}). Device 1 (2) is fabricated with an overlapping gate structure on an undoped Si/SiGe heterostructure with an $8$-nm-wide natural Si quantum well approximately $50~\si{nm}$ below the surface and with a $2~(4)$-nm-thick Si cap layer. In both devices, a $15~\si{nm}$-thick $\text{Al}_2\text{O}_3$ gate dielectric is grown on top of the semiconductors using atomic layer deposition, and the charge sensors are configured for rf reflectometry \cite{PhysRevApplied.13.024019}.  

We tune the upper right sensor quantum dot of Device 1 in the Coulomb blockade regime and set the plunger gate voltage on the side of a transport peak, such that the dot conductance variations measured via rf reflectometry reflect chemical potential fluctuations. The upper left charge sensor is tuned not as a dot but as a channel with a high conductance.
The idling voltages of the main-side accumulation gates and finger gates exceed the threshold for electron accumulation, but the channel is depleted by applying low voltages to the screening gates S and SQ (Fig.~\ref{fig:setup}a). 

We observe pronounced random telegraph noise on the upper right sensor dot, and the TLF switches randomly between the two states labeled as ``0'' and ``1'' (Fig.~\ref{fig:setup}b), causing about $3~\si{\mu eV}$ electrochemical potential fluctuations. We calculate the electrochemical potential fluctuations $\delta \epsilon$ from the variations in the reflectometry signal $\delta V_\text{rf}$ using the equation $\delta \epsilon = \alpha \delta V_\text{rf} /(dV_\text{rf}/dV_\text{P})$ \cite{connors2019low}, with $\alpha=0.091~\si{eV/V}$ the lever arm measured from a Coulomb diamond experiment and $dV_\text{rf}/dV_\text{P}$ the sensor sensitivity at the plunger-gate voltage configuration in the experiment.
The TLF measured here is not the one reported in the same device before \cite{ye2024characterization}. 
We identify the 0 (1) state as the ground (excited) state from measurements of the TLF occupation ``bias''. Regardless of the microscopic model, assuming the TLF is in thermal equilibrium with a reservoir at temperature $T$, the occupation bias $B$ between the 1 state and the 0 state of the TLF obeys $B \equiv N_1/N_0 = \exp(-\Delta E/k_B T)$ \cite{ye2024characterization},
where $N_i$ is the population of the state $i$ with $i=0,1$, $\Delta E>0$ is the energy difference or asymmetry between the 1 state and the 0 state, and $k_B$ is Boltzmann's constant.  To illustrate the temperature sensitivity of the TLF, we plot the switching time $\tau$ and occupation bias $B$ at different mixing chamber temperature $T_\text{MC}$ in Fig.~\ref{fig:setup}c. At high temperature, the TLF switching time decreases and the bias increases, as expected. 

The measurements we report below involve increased TLF temperatures due to gate voltage pulses. It could be the case that in the presence of gate voltage pulses, the TLF is in a steady-state but not in equilibrium with a reservoir at temperature $T$. In this case, we would associate the bias with an effective TLF temperature, which is defined by the value of the bias.

\section{Pulse heating}
\begin{figure}[t!]
{\includegraphics[width= 0.49\textwidth]{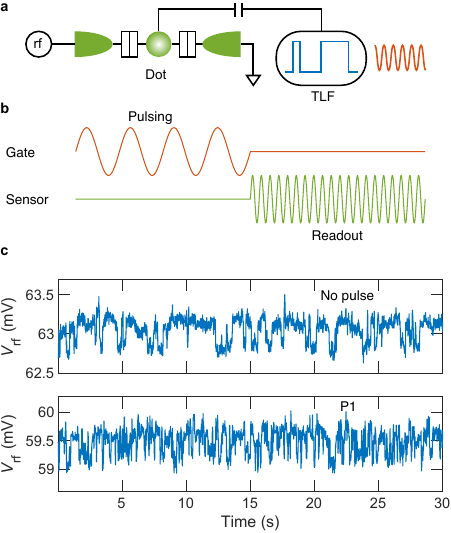}}
\caption{
\textbf{Measurement scheme for pulse heating.}
\textbf{a} The sensor dot measures the TLF, which is heated up by voltage pulses.
\textbf{b} Pulse sequence. We pulse the gate to generate heating and then send an rf excitation to the sensor dot for readout. The total pulse sequence, consisting of the excitation and readout segments, lasts for $20~\si{\mu s}$.
\textbf{c} Example time traces with different pulse configurations. The voltage pulse on the P1 gate has amplitude $10~\si{mV}$ and frequency $2~\si{MHz}$.
}
\label{fig:heating}
\end{figure}

Taking advantage of their strong temperature sensitivity, we use individual charged TLFs probed by the sensor dot to measure pulse heating (Fig.~\ref{fig:heating}a). Unless noted otherwise, the experiments reported here on two devices are conducted at a base temperature of $10~\si{mK}$.
We apply a sinusoidal voltage pulse on a gate to generate heat, and then send an rf excitation with frequency around $200~\si{MHz}$ to the sensor dot through the Ohmic contact for readout (Fig.~\ref{fig:heating}b). This interleaved sequence minimizes spurious effects, including the possibility that the gate voltage pulse changes the sensor-dot conductance, and thus its self heating during excitation~\cite{ye2024characterization}.
We measure the sensor-dot conductance via rf reflectometry \cite{PhysRevApplied.13.024019}, sample the down-converted signal at a $60~\si{Hz}$ rate, and use the averaged signal with the sensor rf excitation on and off as the measured signal $V_\text{rf}$.

Figure~\ref{fig:heating}c shows example time-series for different voltage pulse configurations in Device 1. We find that the average TLF switching time decreases with a voltage pulse applied to the P1 gate, compared to the case without a voltage pulse, suggesting that the TLF is susceptible to pulse heating.

Because we do not have an absolute temperature calibration of the TLF temperature $T$ (partly due to imperfect thermalization of the device to the mixing chamber of the dilution refrigerator), we quantify the pulse heating with the temperature ratio $R\equiv T/T_0$ using measurements of the bias $B$ and the relation $T \propto -\Delta E/\ln B$. Here $T$ is the TLF temperature with a specific heating pulse configuration, and $T_0$ is the TLF temperature with no heating pulse sequence. We assume that the TLF energy difference or asymmetry $\Delta E$ does not depend on the voltage pulse~\cite{niepce2021stability}. We corroborate this assumption by measuring the voltage sensitivity of the TLF occupation bias to each finger gate on the main side (see Supplemental Figure S2 \cite{SMnote}).
To estimate the temperature ratio for each pulse configuration, we measure the TLF without the heating pulse sequence for about 1 minute, and then we measure it again with the heating pulse sequence for about 1 minute. We repeat this pulse-off/pulse-on pair at least 15 times. The entire measurement lasts at least 34 minutes for each pulse configuration. Example traces from a pulse-off/pulse-on pair are shown in Supplemental Figure S1 \cite{SMnote}.

Signal drift and switches in the tuning during the measurement complicate the analysis of the bias, so we post-select for repetitions that do not have significant switches or drift. Specifically, we reject repetitions where the signal range exceeds five times the separation between the two Gaussian peaks with two independent peak widths. This scenario will occur, for example, if the sensor tuning shifts suddenly (see Supplemental Figure S1 \cite{SMnote}). For each of the post-selected repetitions $j$, we compute the number of times we find the TLF in each state $N_{i,j}~(i=0,1)$ as the area under the two Gaussians. We then sum the  populations $N_{i,j}$ across all sets $J_m$ of contiguous repetitions with more than two repetitions per set to calculate a bias for each set $B_m = \sum_{j\in J_m} N_{1,j}/\sum_{j\in J_m} N_{0,j}$. Uncertainty ranges for each $B_m$ value are calculated from the $95\%$ confidence bounds of the double-Gaussian fit parameters. By using this scheme, we can obtain the occupation biases with the pulse off $B_m^\text{off}$ and the pulse on $B_m^\text{on}$ and compute a temperature ratio $R_m$ using the relation $R_m = \log(B_m^\text{off})/\log(B_m^\text{on})$. Finally, we average the temperature ratios $R_m$ across all sets to obtain a weighted averaged temperature ratio $R= \sum_m n_m R_m/\sum_m n_m$ with $n_m$ the number of repetitions in each set, assuming the temperature increase from pulse heating is independent of any switches. The uncertainty in our estimate of $R$ is derived from the uncertainty ranges of $B_m^\text{off}$ and $B_m^\text{on}$.

\begin{table}[t]
\caption{
Temperature variations vs. different pulsed finger gates. $10~\si{mV}$ and $2~\si{MHz}$ sinusoidal pulse is applied on each gate. We interleave 25 pairs of pulse-off and pulse-on measurements here. See Fig.~\ref{fig:setup}a for the device layout. All finger gate voltages are way above the accumulation threshold voltage.
}
\centering 
\begin{ruledtabular}
\begin{tabular}{ccccc}
Gate & P1 & P2 & P3 & P4   \\
$R$ & $1.50\pm0.11$ & $1.23\pm0.08$ & $1.30\pm0.07$ & $1.24\pm0.06$ \\ 
\hline
Gate & T12 & T23 & T34 &  \\
$R$ & $1.37\pm0.07$ & $1.16\pm0.06$ & $1.23\pm0.07$ &
\end{tabular}
\label{table:pulse}
\end{ruledtabular}
\end{table}

\begin{figure*}[ht!]
{\includegraphics[width= 1\textwidth]{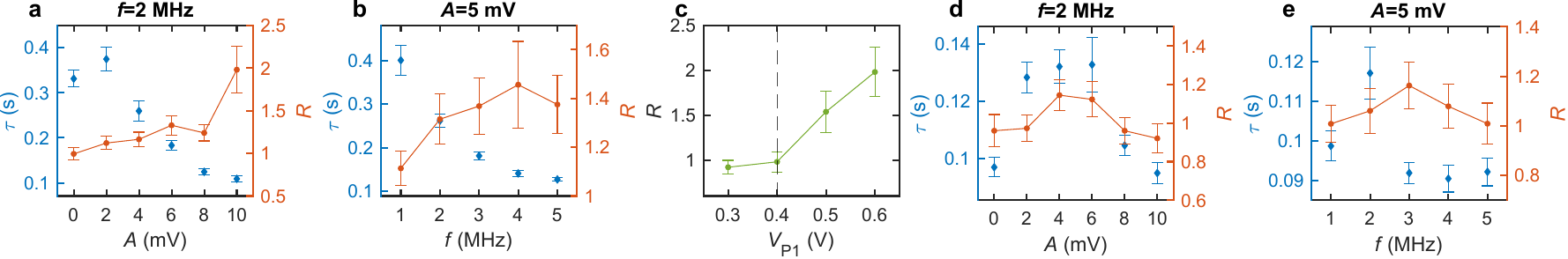}}
\caption{
\textbf{Amplitude, frequency, and voltage dependence of heating from the P1 gate.}
\textbf{a} Switching time and temperature ratio vs. pulse amplitude with $f=2~\si{MHz}$.
\textbf{b} Switching time and temperature ratio vs. pulse frequency with $A=5~\si{mV}$. 
In \textbf{a} and \textbf{b}, $V_\text{P1}=0.6~\si{V}$ is above the electron accumulation threshold of $0.4~\si{V}$.
\textbf{c} Temperature ratio vs. P1 gate idling voltage with a $10~\si{mV}$ and $2~\si{MHz}$ pulse applied to the P1 gate. The vertical dashed line denotes the electron accumulation threshold voltage at $0.4~\si{V}$.
\textbf{d} and \textbf{e} Amplitude and frequency dependence with idling voltage at $0.3~\si{V}$ below the threshold of $0.4~\si{V}$.
}
\label{fig:P1ampfreq}
\end{figure*}

Table~\ref{table:pulse} shows the temperature ratios observed when pulsing different gates with a $10~\si{mV}$ and $2~\si{MHz}$ pulse.
We do not observe a significant correlation between the temperature ratio and the distance from the pulsed gate to the TLF, suggesting that the heating effect is non-local. (Assuming the TLF is localized under the upper-right sensor-dot plunger gate, the closest pulsed finger gate is P4.) Ref.~\cite{PhysRevX.13.041015} also suggests that pulse heating acts globally across a $6$-dot array. 
We emphasize that the main-side ``channel,'' in which qubits would be defined, is depleted by voltages applied to the screening gates S and SQ (Fig.~\ref{fig:setup}a) in these experiments. Thus, the heat source is likely not localized in the main-side channel.

\section{Amplitude and frequency dependence}
To study the amplitude and frequency dependence of pulse heating, we vary the amplitude and frequency of the pulse applied to the P1 gate with the idling voltage $V_\text{P1}$ above the accumulation threshold.
Figures~\ref{fig:P1ampfreq}a-b show that the temperature ratio (switching time) increases (decreases) with the pulse amplitude $A$ and frequency $f$.
We observe a similar amplitude and frequency dependence of other pulsed gates in Device 1 (see Supplemental Figures S3 and S4 \cite{SMnote}) and Device 2 (see Supplemental Figures S8 and S9 \cite{SMnote}).
In this work, we study the effect of MHz-range pulses, which are relevant for singlet-triplet and exchange-only qubits~\cite{burkard2023semiconductor}. Our results indicate that these relatively low-frequency pulses generate heating, consistent with other reports of heating induced by baseband pulses \cite{PhysRevX.13.041015}.

\section{Voltage dependence}
Surprisingly, we find that the temperature ratio depends on the idling voltage of the pulsed gates. Fig.~\ref{fig:P1ampfreq}c illustrates how the temperature ratio for a $10~\si{mV}$ and $2~\si{MHz}$ voltage pulse applied to the P1 gate depends on the idling voltage of that gate.  The data indicate that $R$ approaches $1$ when the P1 idling voltage is below the electron accumulation threshold of $0.4~\si{V}$, suggesting that the heating is mitigated.
At $V_\mathrm{P1}=0.3~\si{V}$, we measure amplitude and frequency dependence of the heating (Figs.~\ref{fig:P1ampfreq}d-e), and find that $R$ fluctuates around 1 as a function of amplitude and frequency, confirming that the pulse heating is mitigated when the idling voltage of the pulsed gate is far below the accumulation threshold.
We also observe voltage-dependent heating for other finger gates in Device 1 (see Supplemental Figure S5 \cite{SMnote}) and Device 2 (see Supplemental Figures S10 and S11 \cite{SMnote}).

\section{Interpretation of pulse heating}
\begin{figure}[b]
{\includegraphics[width= 0.5\textwidth]{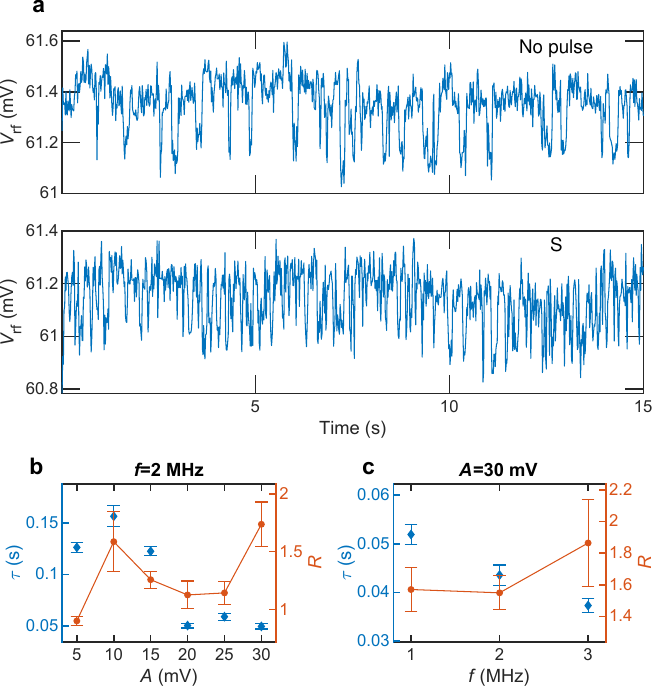}}
\caption{
\textbf{Heating caused by pulsing the middle screening gate S.}
\textbf{a} Example time-series without (top) and with (bottom) a pulse applied to the S gate with $A=30~\si{mV}$ and $f=2~\si{MHz}$.
\textbf{b} Switching time and temperature ratio vs. pulse amplitude with $f=2~\si{MHz}$. 
\textbf{c} Switching time and temperature ratio vs. pulse frequency with $A=30~\si{mV}$.
}
\label{fig:Sgate}
\end{figure}

As shown in Fig.~\ref{fig:Sgate}a, applying a voltage pulse to the middle screening gate S also causes the temperature ratio $R$  (switching time $\tau$) to increase (decrease). While we do not see a monotonic increase as in the case of finger gates (possibly due to limited signal-to-noise ratio and occasional signal switches), we observe that $R$ tends to increase with pulse amplitude and frequency (Figs.~\ref{fig:Sgate}b-c) as with the other finger gates. For this device, the heating associated with the middle screening gate is significant: with a $30~\si{mV}$ and $2~\si{MHz}$ sinusoidal voltage pulse applied to the middle screening gate S, the TLF temperature increases by about $1.6$ times.

Note however that the idling voltage of the S gate is well below the accumulation threshold for electrons. (Without a low voltage applied to this gate, quantum dots cannot easily be formed.) Thus, unlike the situation with the finger gates, applying a voltage pulse to the S gate with a low idling voltage appears to generate heat.  
We can reconcile these observations by hypothesizing that pulsing a gate with electrons nearby or underneath that gate causes a temperature increase. 
For the pulsed finger gates, electrons accumulated in the quantum well in the fanout regions (Fig.~\ref{fig:setup}a) are a possible source of heating. For the middle screening S gate, electrons underneath the nearby accumulation gates could be the origin.

The precise mechanism of the heating remains unclear, however. Joule heating or other scattering processes associated with the electrons in the quantum well are possible causes of the pulse heating we observe, and these would be consistent with the observed dependence on the idling voltage. In this case, thermal conduction between electrons in the quantum well and the TLF via phonons or the Coulomb interaction could increase the temperature of the TLF. However, dielectric loss is another potential cause of pulse heating if we consider that electrons near the gates, together with the gates themselves, could form an effective capacitor. Here, it would seem that thermal conduction via phonons would heat the TLF. 

While we cannot pinpoint the mechanism of pulse heating, we hypothesize that decreasing the area of the gates with electrons nearby could mitigate pulse heating. For the finger gates, decreasing the area where the gates overlap the quantum well would help.
Our hypothesis, which requires further experimental testing, is motivated by the observation that apparently depleting the electrons under the gates reduces the heating.
For the middle screening gate, moving the neighboring accumulation gates away from it could reduce the heating. In addition, we anticipate that moving gate electrodes far away from the heterostructure using vertical vias directly connecting the active gates \cite{Ha2022sledge} could reduce the heating.

\section{Conclusion}
In conclusion, we have measured pulse heating in Si quantum dots using individual charged TLFs. Our work suggests that pulsing the gates with electrons nearby causes dissipation. In addition to the existing methods involving optimizing controlled pulses \cite{takeda2018optimized,zwerver2022qubits,sato2024simulation} and operating devices at elevated temperature \cite{PhysRevX.13.041015}, we hypothesize that reducing the area of the gates near the electrons could mitigate the pulse heating. The connection of this pulse heating to pulse-induced frequency shifts remains an open question, although our work shows that electrical fluctuators are also susceptible to pulse heating, in line with several theoretical works \cite{PhysRevResearch.6.013168,sato2024simulation} that have used randomly distributed TLFs to model pulsed-induced frequency shifts. Our results also motivate future work on using TLFs in semiconductor quantum dots as local thermometers.

\section{Data Availability}
The processed data that support the findings of this study are available in Ref.~\cite{zenodo}. The raw data are available from the corresponding author upon reasonable request.

\section{Acknowledgments}
We thank Lisa F. Edge of HRL Laboratories for growing the heterostructures and Elliot J. Connors for fabricating the devices used in this work.
This work was sponsored by the Army Research Office through Grant No. W911NF-23-1-0115 and the Air Force Office of Scientific Research through Grant No. FA9550-23-1-0710. The views and conclusions contained in this document are those of the authors and should not be interpreted as representing the official policies, either expressed or implied, of the Army Research Office or the U.S. Government. The U.S. Government is authorized to reproduce and distribute reprints for Government purposes notwithstanding any copyright notation herein.

%

\end{document}


\title{Supplemental Material for \\ ``Measuring pulse heating in Si quantum dots with individual two-level fluctuators"}

\author{Feiyang Ye}

\author{Lokendra S. Dhami}

\affiliation{Department of Physics and Astronomy, University of Rochester, Rochester, NY, 14627, USA}
\affiliation{University of Rochester Center for Coherence and Quantum Science, Rochester, NY, 14627, USA}

\author{John M. Nichol}
\email{john.nichol@rochester.edu}
\affiliation{Department of Physics and Astronomy, University of Rochester, Rochester, NY, 14627, USA}
\affiliation{University of Rochester Center for Coherence and Quantum Science, Rochester, NY, 14627, USA}

\maketitle

\tableofcontents

\clearpage
\section{Supplemental data for Device 1}


\subsection{Post-selection scheme}

\begin{figure}[ht]
\centering
{\includegraphics[width=0.9 \textwidth]{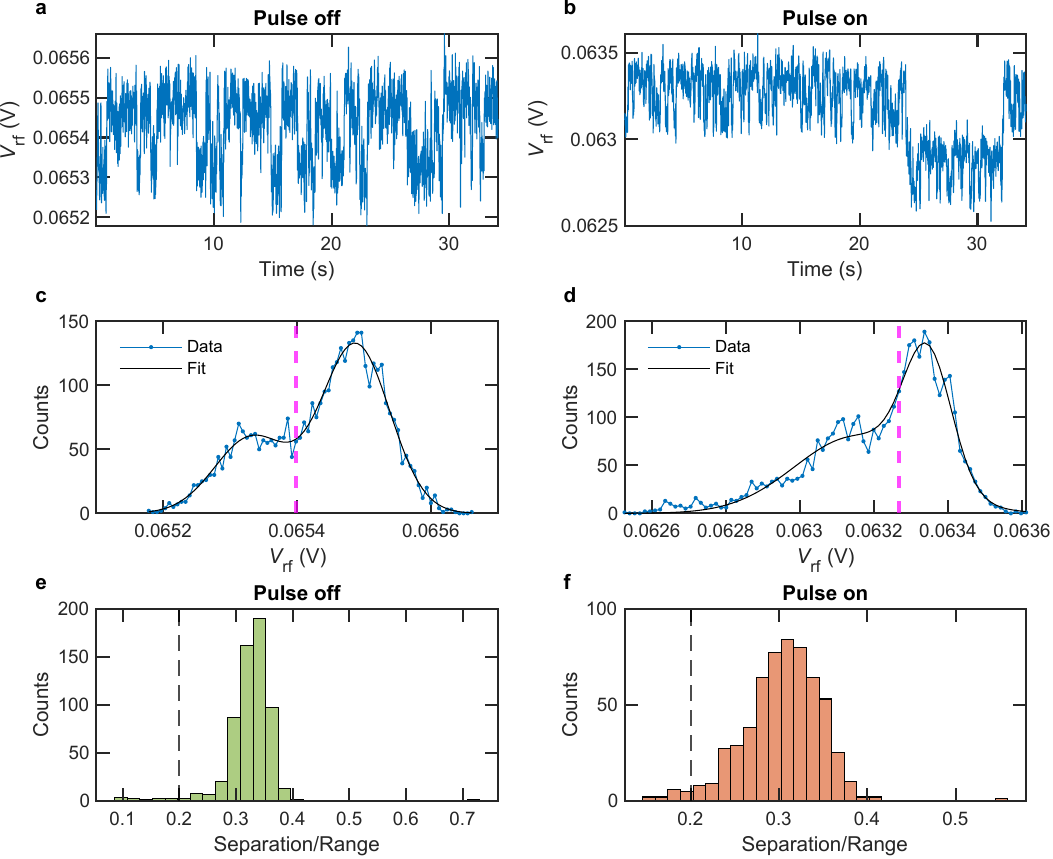}}
\caption{
\textbf{Post-selection scheme.}
\textbf{a} Example time-series when there is no pulse applied to the P1 gate.
\textbf{b} Example time-series when there is a $5~\si{mV}$ and $4~\si{MHz}$ pulse applied to the P1 gate.
Panels \textbf{a} and \textbf{b} show examples of a pair of alternate pulse-off and pulse-on scans measured in Fig.~3b in the main text.
\textbf{c} Signal histogram from \textbf{a} and fit to the sum of two Gaussians.
\textbf{d} Signal histogram from \textbf{b} and fit to the sum of two Gaussians. The signal switch changes the signal range, and the separation of the two Gaussian peaks is about 18\% of the range of the signals, which is below the 20\% threshold. Hence, the post-selection scheme removes this pair of time traces from the analysis.
\textbf{e} Histogram of separation/range for the pulse-off data in Figs. 3 and 4 in the main text.
\textbf{f} Histogram of separation/range for the pulse-on data in Figs. 3 and 4 in the main text.
In \textbf{e} and \textbf{f}, the 20\% post-selection threshold is indicated by a vertical line.
}
\label{figS:post-sel}
\end{figure}

\begin{table}[h!]
\centering
\caption{$J_m$ and $n_m$ values for Figures 3a–3c. We omit the data points where repetitions are included by the post-selection scheme.}
\renewcommand{\arraystretch}{1.2}
\begin{tabular}{|c|c|c|c|}
\hline
\textbf{Figure} & \textbf{Parameter} & \textbf{$J_m$} & \textbf{$n_m$} \\ \hline

\multirow{1}{*}{3a} 
 & $A$ = 10 mV & [1--3], [5--10] & [3], [6] \\ \hline

\multirow{4}{*}{3b} 
 & $f$ = 2 MHz & [1--3], [5--15] & [3], [11] \\ \cline{2-4}
 & $f$ = 3 MHz & [1--5], [7--15] & [5], [9] \\ \cline{2-4}
 & $f$ = 4 MHz & [1--3], [8--11], [13--15] & [3], [4], [3] \\ \cline{2-4}
 & $f$ = 5 MHz & [1--12] & [12] \\ \hline

\multirow{4}{*}{3c}
 & $V_\text{P1}$ = 0.3 V & [3--14] & [12] \\ \cline{2-4}
 & $V_\text{P1}$ = 0.4 V & [7--14] & [8] \\ \cline{2-4}
 & $V_\text{P1}$ = 0.5 V & [1--6], [10--12] & [6], [3] \\ \cline{2-4}
 & $V_\text{P1}$ = 0.6 V & [1--3], [5--10] & [3], [6] \\ \hline

\end{tabular}
\end{table}

\begin{table}[h!]
\centering
\caption{$J_m$ and $n_m$ values for Figures 3d–3e.}
\renewcommand{\arraystretch}{1.2}
\begin{tabular}{|c|c|c|c|}
\hline
\textbf{Figure} & \textbf{Parameter} & \textbf{$J_m$} & \textbf{$n_m$} \\ \hline

\multirow{5}{*}{3d} 
 & $A$ = 0 mV & [2--7], [10--15] & [6], [6] \\ \cline{2-4}
 & $A$ = 2 mV & [3--15] & [13] \\ \cline{2-4}
 & $A$ = 4 mV & [2--15] & [14] \\ \cline{2-4}
 & $A$ = 6 mV & [1--6], [8--10], [12--15] & [6], [3], [4] \\ \cline{2-4}
 & $A$ = 10 mV & [3--14] & [12] \\ \hline

\multirow{3}{*}{3e}
 & $f$ = 2 MHz & [1--5], [7--15] & [5], [9] \\ \cline{2-4}
 & $f$ = 3 MHz & [3--15] & [13] \\ \cline{2-4}
 & $f$ = 5 MHz & [1--4], [6--10], [12--15] & [4], [5], [4] \\ \hline

\end{tabular}
\end{table}

\begin{table}[h!]
\centering
\caption{$J_m$ and $n_m$ values for Figure 4.}
\begin{tabular}{|c|c|c|c|}
\hline
\textbf{Figure} & \textbf{Parameter} & \textbf{$J_m$} & \textbf{$n_m$} \\ \hline
\multirow{5}{*}{4b} 
 & $A$ = 10 mV & [6--8], [21--25] & [3], [5] \\ \cline{2-4}
 & $A$ = 15 mV & [1--4], [9--11], [13--15], [17--20], [23--25] & [4], [3], [3], [4], [3] \\ \cline{2-4}
 & $A$ = 20 mV & [3--5], [10--12], [14--19], [23--25] & [3], [3], [6], [3] \\ \cline{2-4}
 & $A$ = 25 mV & [3--7], [9--14], [18--20] & [5], [6], [3] \\ \cline{2-4}
 & $A$ = 30 mV & [1--6], [8--10], [12--16], [18--25] & [6], [3], [5], [8] \\ \hline
\multirow{3}{*}{4c} 
 & $f$ = 1 MHz & [1--7], [14--24] & [7], [11] \\ \cline{2-4}
 & $f$ = 2 MHz & [1--20], [22--25] & [20], [4] \\ \cline{2-4}
 & $f$ = 3 MHz & [1--4], [14--17], [20--25] & [4], [4], [6] \\ \hline
\end{tabular}
\end{table}

\clearpage
\subsection{Voltage sensitivity of the TLF}
\begin{figure}[ht]
\centering
{\includegraphics[width=1 \textwidth]{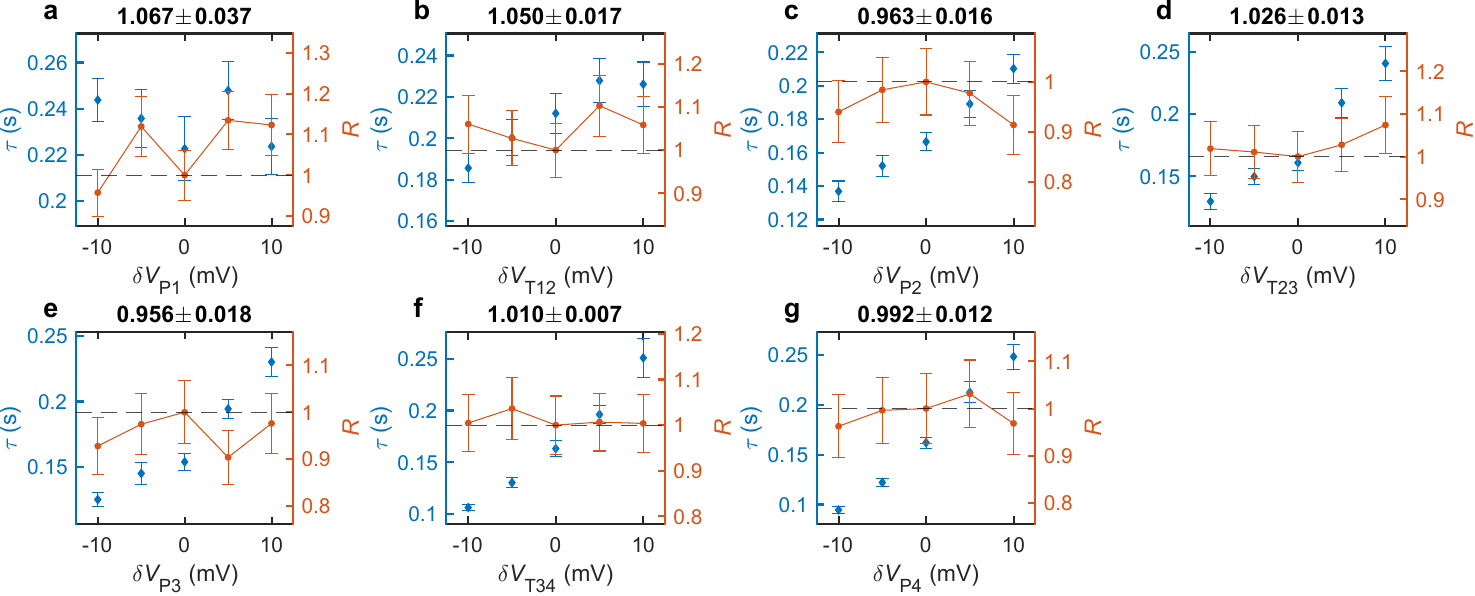}}
\caption{
\textbf{Voltage sensitivity of the TLF in Device 1.}
Switching time and temperature ratio vs. gate voltage variation. All gate voltages are far above the threshold voltage and are swept $\pm 10~\si{mV}$ from the idling voltage. In the main text, the maximum amplitude of voltage pulses applied to the finger gates in Device 1 is $10~\si{mV}$. 
}
\label{figS:gateD1}
\end{figure}

To quantify the voltage sensitivity of the TLF, we sweep the DC voltage bias $\pm 10~\si{mV}$ from the idling voltage of each finger gate, and measure how the TLF responds to the gate voltage variation. 
Figure~\ref{figS:gateD1} plots the TLF switching time $\tau$ and temperature ratio $R \equiv T/T_0$ as a function of gate voltage variation $\delta V$ for each finger gate on the main side. We use the data at the idling voltage tuning to identify the TLF temperature baseline $T_0$. 
The title of each subplot shows that the average temperature ratio is close to $1$ within the $\pm 10~\si{mV}$ voltage variations, indicating that all finger gates couple weakly to the TLF.
The systematic trend in the switching time vs. gate voltage is due to the self-heating caused by the change in the sensor-dot conductance \cite{refA} arising from the voltage crosstalk effect between the sensor dot and the finger gates on the main side.

\clearpage
\subsection{Pulse amplitude and frequency dependence}
\begin{figure}[ht]
\centering
{\includegraphics[width=1 \textwidth]{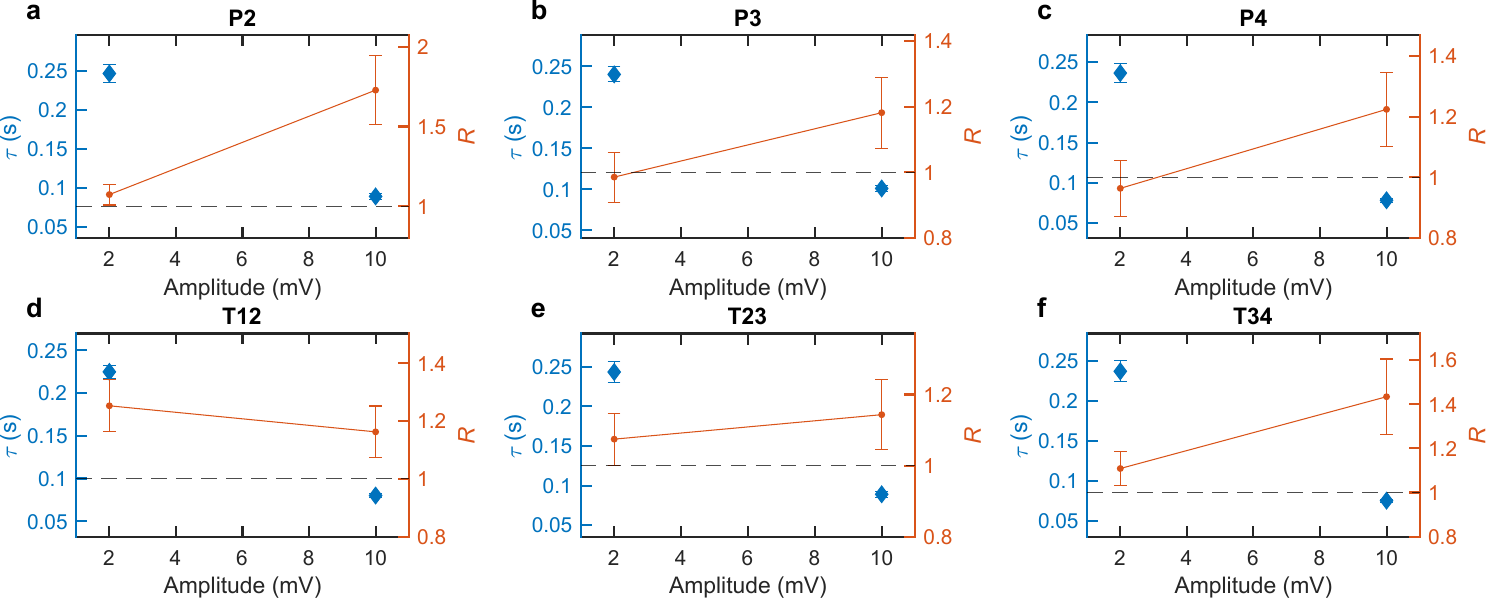}}
\caption{
\textbf{Pulse amplitude dependence.} 
The switching time (temperature ratio) tends to decrease (increase) with amplitude. The pulse frequency is $2~\si{MHz}$.
The title of each plot shows the pulsed gate. All finger gate idling voltages are a few hundred millivolts above the threshold voltage of $400~\si{mV}$. The horizontal dashed line denotes $R=1$.
}
\label{figS:ampD1}
\end{figure}

\begin{figure}[ht]
\centering
{\includegraphics[width=1 \textwidth]{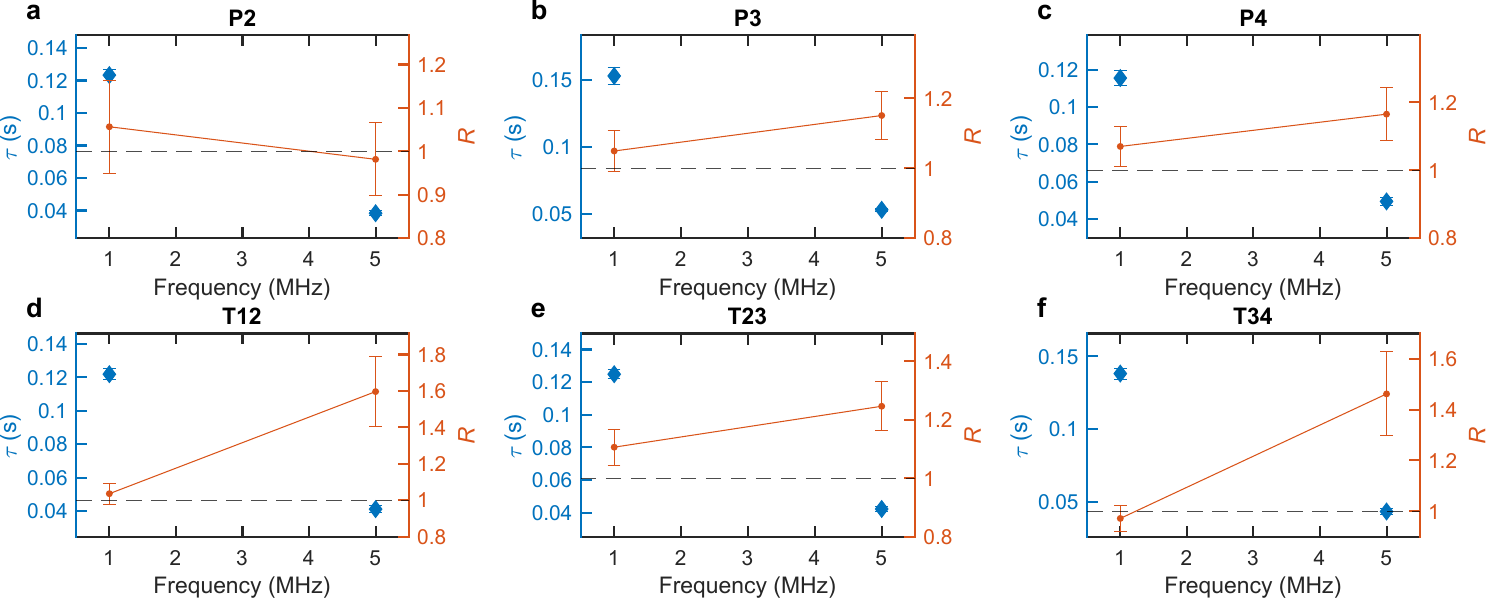}}
\caption{
\textbf{Pulse frequency dependence.}
The switching time (temperature ratio) tends to decrease (increase) with frequency. The pulse amplitude is $5~\si{mV}$.
}
\label{figS:freqD1}
\end{figure}

\clearpage

\begin{figure}[ht]
\centering
{\includegraphics[width=1 \textwidth]{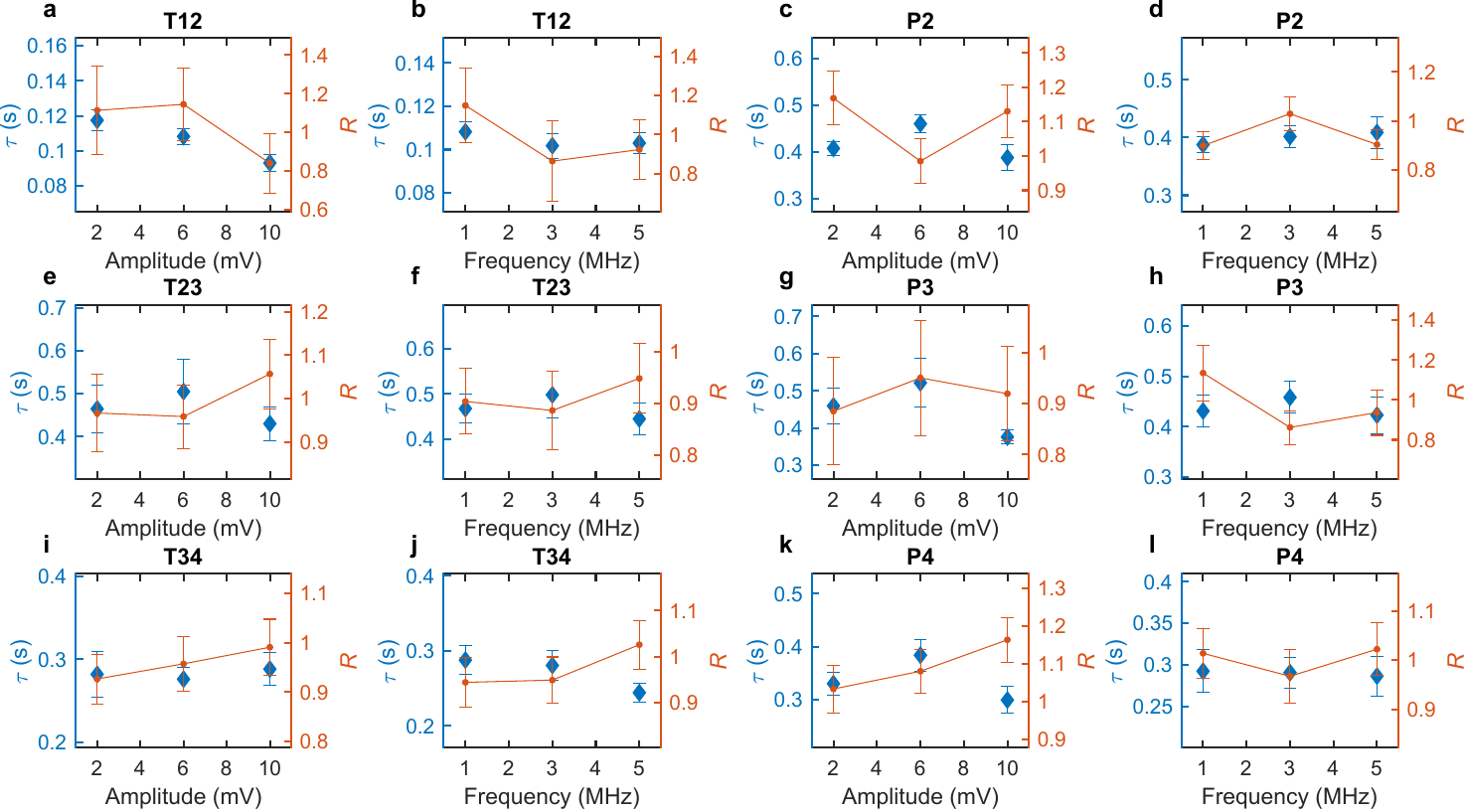}}
\caption{
\textbf{Pulse amplitude and frequency dependence with idling voltage below the threshold.}
The title of each plot indicates the pulsed gate.
In contrast to Figs.~\ref{figS:ampD1} and \ref{figS:freqD1}, there is no amplitude and frequency dependence when the idling voltage $0.25~\si{V}$ is below the accumulation threshold of $0.4~\si{V}$. The temperature ratio $R$ fluctuates around $1$, indicating that the pulse heating is mitigated.
}
\label{figS:voltD1}
\end{figure}

\clearpage
\section{Supplemental data for Device 2}

\subsection{Experimental setup of Device 2}
\begin{figure}[b]
{\includegraphics[width= 0.55\textwidth,trim = 6 7 7 6,clip]{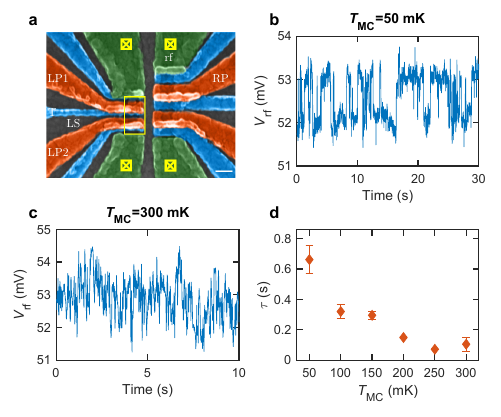}}
\caption{
\textbf{Experimental setup of Device 2.}
\textbf{a} Scanning electron micrograph of a double-quantum-dot device. The sensor quantum dot is tuned underneath plunger gate RP and is configured for rf reflectometry. The left main-side channel is depleted by applying low voltages to the screening gate LS and middle screening gate. Voltage pulses are sent through three finger gates LP1, T and LP2. The wiring of the middle screening gate in Device 2 is not configured for pulsing. The white scale bar is $100~\si{nm}$.
\textbf{b} Example time-series at a mixing chamber temperature of $50~\si{mK}$.
\textbf{c} Example time-series at a mixing chamber temperature of $300~\si{mK}$.
\textbf{d} Switching time vs. mixing chamber temperature. 
In \textbf{b}-\textbf{d}, the sensor rf excitation is turned on half of the time as described in the main text. 
}
\label{figS:JN49}
\end{figure}

We measure Device 2 (Fig.~\ref{figS:JN49}a) in a different but nominally similar dilution refrigerator.
The sensor dot is tuned under plunger gate RP in the Coulomb blockade regime. The left main-side tuning is similar to that of Device 1, with accumulation-gate and finger-gate voltages tuned above the accumulation threshold while the channel is pinched off by applying low voltages on the middle screening gate and screening gate LS.
We observe another large-amplitude TLF in Device 2 (Fig.~\ref{figS:JN49}b). To measure the temperature dependence of the TLF, we sweep the mixing chamber temperature from $50~\si{mK}$ to $300~\si{mK}$ with a step of $50~\si{mK}$, and take a 32-minute-long time-series at each temperature. Examples of the time traces at different temperatures (Figs.~\ref{figS:JN49}b,c) indicate that the switching between the two states is more frequent at higher temperature. The switching time of the TLF in Device 2 tends to decrease with temperature (Fig.~\ref{figS:JN49}d). 
The relatively low signal to noise ratio in Device 2 prevents us from extracting the occupation bias.

\subsection{Voltage sensitivity of the TLF}
\begin{figure}[ht]
\centering
{\includegraphics[width=0.8 \textwidth]{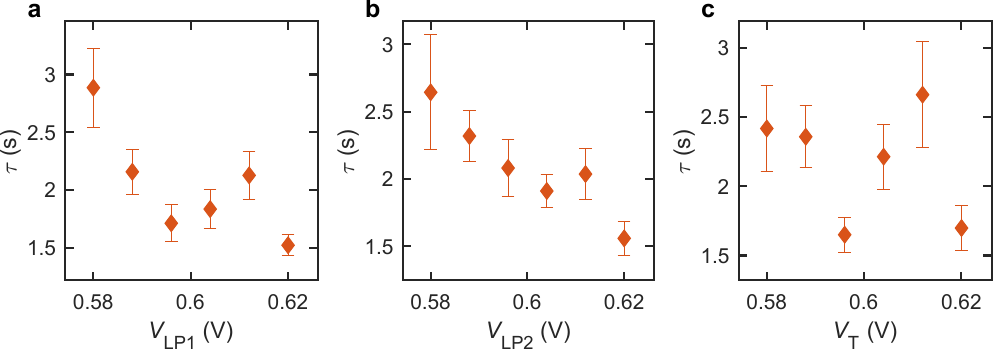}}
\caption{
\textbf{Voltage sensitivity of the TLF in Device 2.}
TLF switching time vs. gate voltage. The idling voltage of each finger gate is tuned at $600~\si{mV}$. The DC voltage bias is swept $\pm 20~\si{mV}$ from the idle tuning. The maximum amplitude of voltage pulses applied in Device 2 is $20~\si{mV}$.
In \textbf{a} and \textbf{b}, the switching time tends to decrease with plunger gate voltage because the sensor-dot conductance increases due to capacitive coupling to the plunger gates. This increased conductance leads to a heating effect caused by the current through the dot \cite{refA}, which in turn increases the temperature of the TLF. 
In \textbf{c}, there is no clear trend between the switching time and the tunneling gate voltage.
}
\label{figS:gateD2}
\end{figure}
\clearpage

\subsection{Pulse amplitude and frequency dependence}

\begin{figure}[ht]
\centering
{\includegraphics[width=0.8 \textwidth]{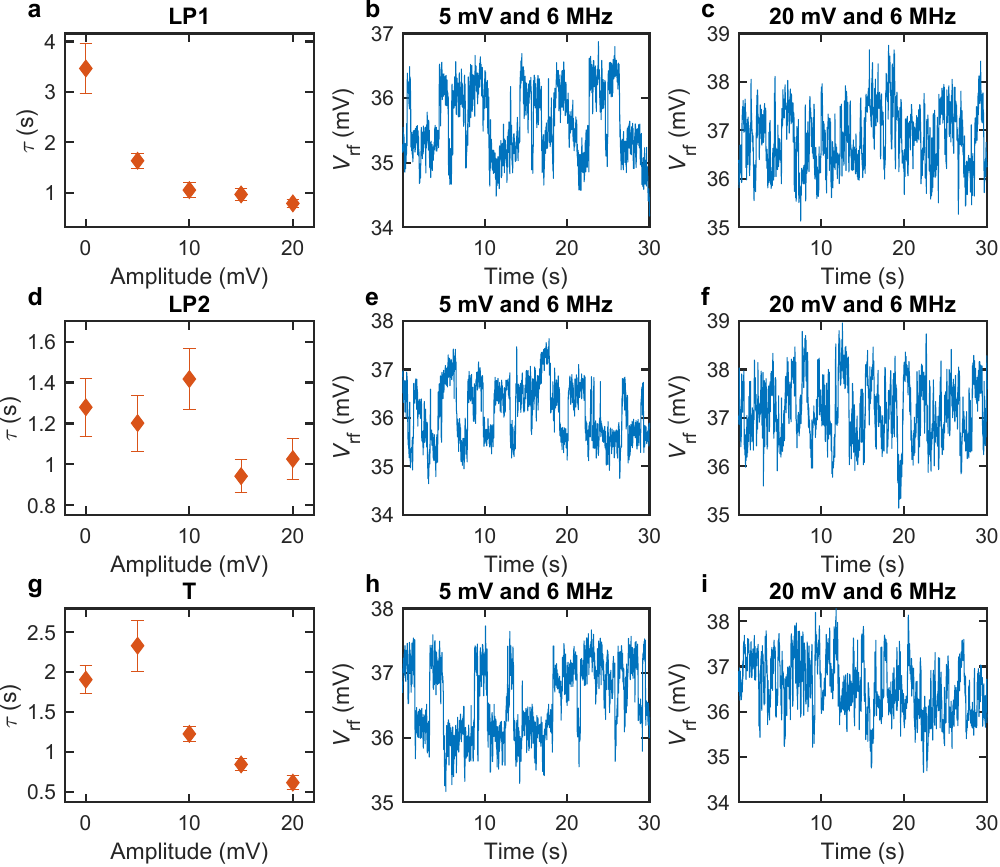}}
\caption{
\textbf{Pulse amplitude dependence with idling voltage above the threshold.}
Idling gate voltages are tuned at $0.6~\si{V}$ above the accumulation threshold of $0.3~\si{V}$ in Device 2. Pulse frequency is $6~\si{MHz}$.
\textbf{a}-\textbf{c} Pulse amplitude dependence for the plunger gate LP1. The TLF temperature increases with pulse amplitude, indicated by a decreasing switching time. However, the voltage-induced effect observed in Fig.~\ref{figS:gateD2} could be the cause of the temperature increase, instead of pulse heating.
\textbf{d}-\textbf{f} Pulse amplitude dependence for the plunger gate LP2.
\textbf{g}-\textbf{i} Pulse amplitude dependence for the tunneling gate T. 
}
\label{figS:ampD2}
\end{figure}

\begin{figure}[ht]
\centering
{\includegraphics[width=0.8 \textwidth]{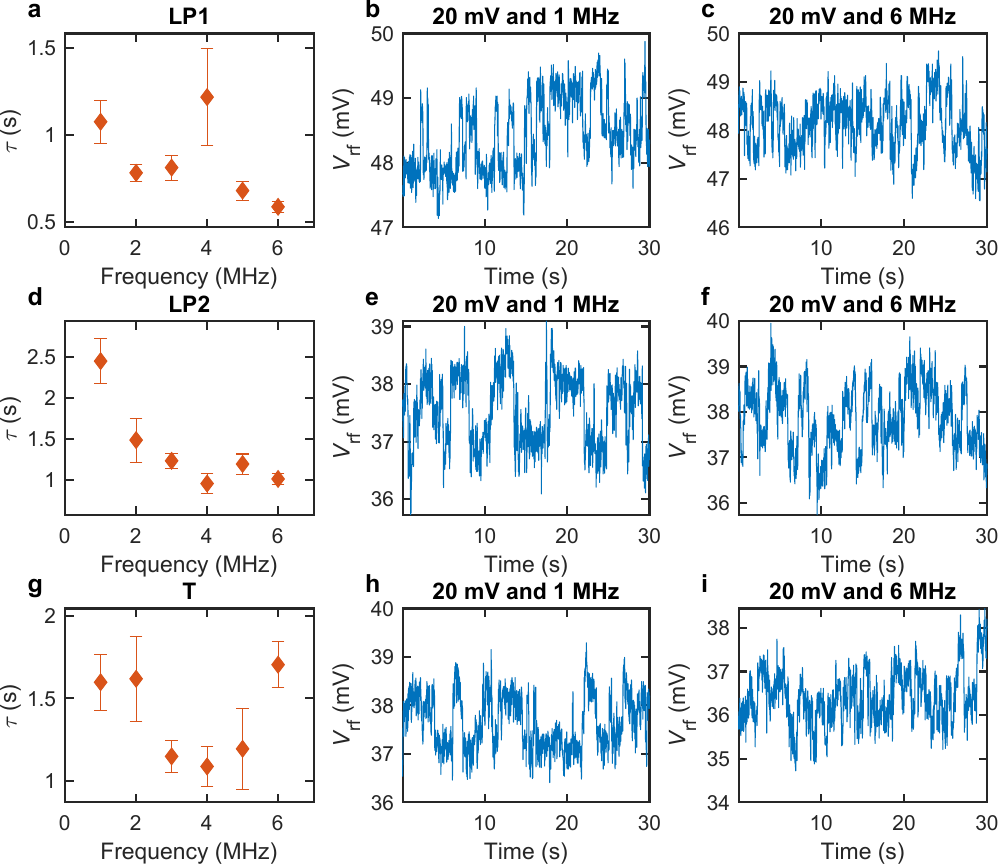}}
\caption{
\textbf{Pulse frequency dependence with idling voltage above the threshold.}
Pulse amplitude is $20~\si{mV}$.
\textbf{a}-\textbf{c} Pulse frequency dependence for the plunger gate LP1.
\textbf{d}-\textbf{f} Pulse frequency dependence for the plunger gate LP2.
\textbf{g}-\textbf{i} Pulse frequency dependence for the tunneling gate T. 
}
\label{figS:freqD2}
\end{figure}

\clearpage
\subsection{Voltage-dependent pulse heating}

\begin{figure}[ht]
\centering
{\includegraphics[width=0.8 \textwidth]{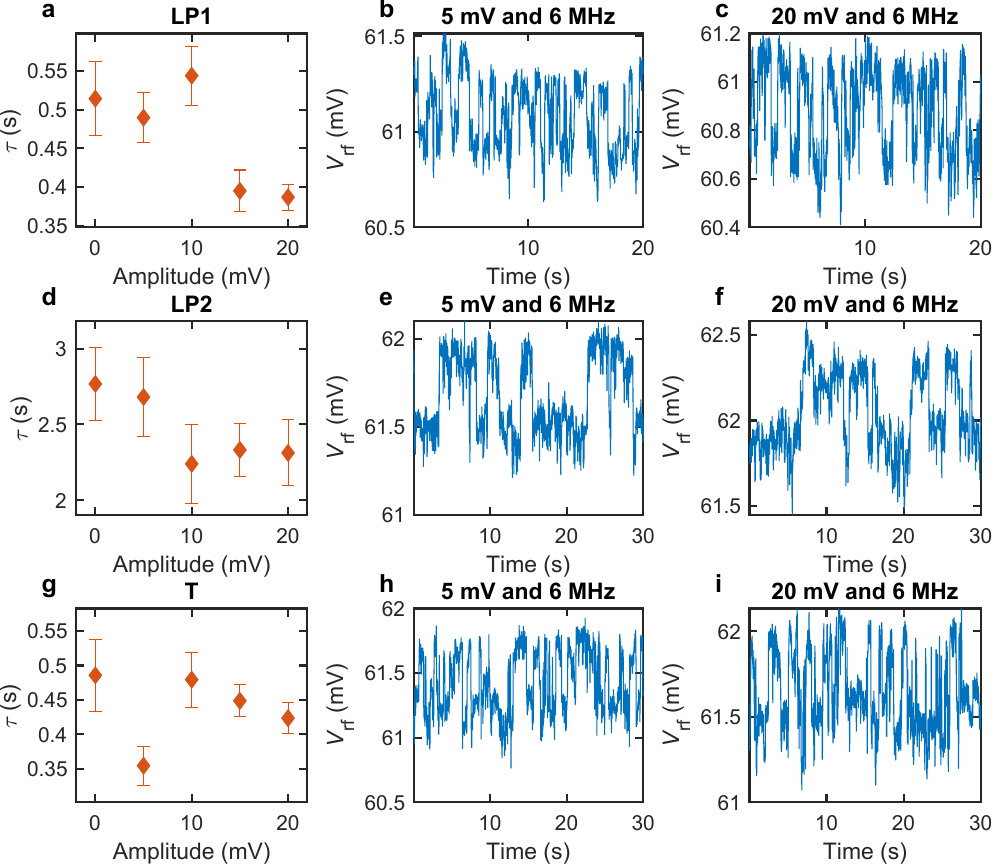}}
\caption{
\textbf{Pulse amplitude dependence with idling voltage below the threshold.} 
Idling gate voltages are tuned at $0.24~\si{V}$ below the accumulation threshold of $0.3~\si{V}$.
\textbf{a}-\textbf{c} Pulse amplitude dependence for the plunger gate LP1.
\textbf{d}-\textbf{f} Pulse amplitude dependence for the plunger gate LP2.
\textbf{g}-\textbf{i} Pulse amplitude dependence for the tunneling gate T.
In contrast to Fig.~\ref{figS:ampD2}, pulse heating is mitigated when the idling voltage of the pulsed gate is below the accumulation threshold.
}
\label{figS:volt_amp_D2}
\end{figure}

\begin{figure}[ht]
\centering
{\includegraphics[width=0.8 \textwidth]{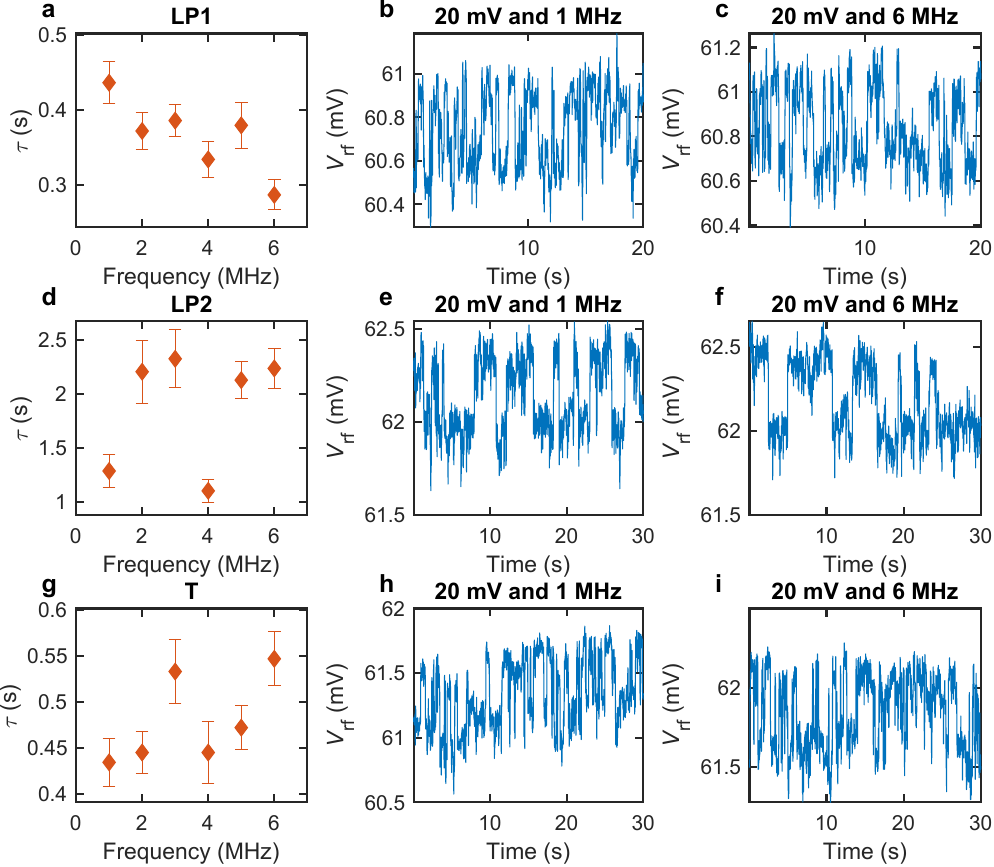}}
\caption{
\textbf{Pulse frequency dependence with idling voltage below the threshold.}
Idling gate voltages are tuned at $0.24~\si{V}$ below the accumulation threshold of $0.3~\si{V}$.
\textbf{a}-\textbf{c} Pulse frequency dependence for the plunger gate LP1.
\textbf{d}-\textbf{f} Pulse frequency dependence for the plunger gate LP2.
\textbf{g}-\textbf{i} Pulse frequency dependence for the tunneling gate T.
Compared to Fig.~\ref{figS:freqD2}, pulse heating is suppressed when the electrons are depleted under the pulsed gate.
}
\label{figS:volt_freq_D2}
\end{figure}

\newpage

